\documentclass[sigconf,screen]{acmart}

\usepackage[normalem]{ulem}
\usepackage{soul}
\usepackage{amsmath}

\usepackage{amssymb}
\usepackage{amsfonts}
\usepackage{algorithmic}
\usepackage{graphicx}
\usepackage{enumitem}

\usepackage{pythonhighlight}
\usepackage{multirow}

\definecolor{codegreen}{rgb}{0,0.6,0}
\definecolor{codegray}{rgb}{0.5,0.5,0.5}
\definecolor{codepurple}{rgb}{0.58,0,0.82}
\definecolor{backcolour}{rgb}{0.95,0.95,0.92}

\lstdefinestyle{mystyle}{
    backgroundcolor=\color{backcolour},   
    commentstyle=\color{codegreen},
    keywordstyle=\color{magenta},
    numberstyle=\tiny\color{codegray},
    stringstyle=\color{codepurple},
    basicstyle=\ttfamily\footnotesize,
    breakatwhitespace=false,         
    breaklines=true,                 
    captionpos=b,                    
    keepspaces=true,                 
    numbers=left,                    
    numbersep=5pt,                  
    showspaces=false,                
    showstringspaces=false,
    showtabs=false,                  
    tabsize=2
}

\usepackage{hyperref}

\usepackage{pifont}

\copyrightyear{2023}
\acmYear{2023}
\setcopyright{acmlicensed}\acmConference[ISCA '23]{Proceedings of the 50th Annual International Symposium on Computer Architecture}{June 17--21, 2023}{Orlando, FL, USA}
\acmBooktitle{Proceedings of the 50th Annual International Symposium on Computer Architecture (ISCA '23), June 17--21, 2023, Orlando, FL, USA}
\acmPrice{15.00}
\acmDOI{10.1145/3579371.3589072}
\acmISBN{979-8-4007-0095-8/23/06}

\begin{document}
\begin{sloppypar}

\title{Mystique: Enabling Accurate and Scalable Generation of Production AI Benchmarks}

\author{Mingyu Liang}
\affiliation{%
	\institution{Cornell University}
	\city{Ithaca}
	\state{New York}
	\country{USA}
}
\email{ml2585@cornell.edu}

\author{Wenyin Fu}
\affiliation{%
	\institution{Meta}
	\city{Menlo Park}
	\state{California}
	\country{USA}
}
\email{wenyinfu@meta.com}

\author{Louis Feng}
\affiliation{%
	\institution{Meta}
	\city{Menlo Park}
	\state{California}
	\country{USA}
}
\email{lofe@meta.com}

\author{Zhongyi Lin}
\affiliation{%
	\institution{University of California, Davis}
	\city{Davis}
	\state{California}
	\country{USA}
}
\email{zhylin@ucdavis.edu}

\author{Pavani Panakanti}
\affiliation{%
	\institution{Meta}
	\city{Menlo Park}
	\state{California}
	\country{USA}
}
\email{pavanip@meta.com}

\author{Shengbao Zheng}
\affiliation{%
	\institution{Meta}
	\city{Menlo Park}
	\state{California}
	\country{USA}
}
\email{shengbao@meta.com}

\author{Srinivas Sridharan}
\affiliation{%
	\institution{Meta}
	\city{Menlo Park}
	\state{California}
	\country{USA}
}
\email{ssrinivas@meta.com}

\author{Christina Delimitrou}
\affiliation{%
	\institution{Massachusetts Institute of Technology}
	\city{Cambridge}
	\state{Massachusetts}
	\country{USA}
}
\email{delimitrou@csail.mit.edu}

\renewcommand{\shortauthors}{M. Liang, W. Fu, L. Feng, Z. Lin, P. Panakanti, S. Zheng, S. Sridharan, and C. Delimitrou}

\ccsdesc[500]{Computing methodologies~Artificial intelligence}
\ccsdesc[500]{Computer systems organization~Cloud computing}
\ccsdesc[500]{Computer systems organization~Neural networks}
\ccsdesc[300]{Software and its engineering~Source code generation}

\keywords{artificial intelligence, cloud computing, benchmarking, performance cloning, code generation}

\begin{abstract} 

Building large AI fleets to support the rapidly growing DL workloads is an active research topic for modern cloud providers. Generating accurate benchmarks plays an essential role in designing the fast-paced software and hardware solutions in this space. Two fundamental challenges to make this scalable are (i) workload representativeness and (ii) the ability to quickly incorporate changes to the fleet into the benchmarks. 

To overcome these issues, we propose Mystique, an accurate and scalable framework for production AI benchmark generation. It leverages the PyTorch execution trace (ET), a new feature that captures the runtime information of AI models at the granularity of operators, in a graph format, together with their metadata. By sourcing fleet ETs, we can build AI benchmarks that are portable and representative. Mystique is scalable, due to its lightweight data collection, in terms of runtime overhead and instrumentation effort. It is also adaptive because ET composability allows flexible control on benchmark creation. 

We evaluate our methodology on several production AI models, and show that benchmarks generated with Mystique closely resemble original AI models, both in execution time and system-level metrics. We also showcase the portability of the generated benchmarks across platforms, and demonstrate several use cases enabled by the fine-grained composability of the execution trace. 
\end{abstract}

\maketitle

\section{Introduction}

Artificial Intelligence (AI) has experienced a strong resurgence with the recent advances in Deep Learning (DL). It is rapidly expanding into many areas, and has led to revolutionary changes, including in natural language processing~\cite{devlin2018bert, brown2020language}, computer vision~\cite{he2016deep, szegedy2015going}, gaming~\cite{silver2016mastering, vinyals2019grandmaster}, and recommendation systems~\cite{naumov2019deep, zhou2018deep}. Almost all cloud enterprises today deploy massive amounts of resources towards AI computing to support their business. Building and maintaining large AI fleets to efficiently support these DL workloads has led to both hardware and software innovation across the system stack~\cite{zhao22,delimitrou14,chen2014diannao,gan18,gan19,gan21, jouppi2017datacenter, norrie2021design}. 

Having representative and agile AI benchmarks based on live fleet production workloads would provide an invaluable resource for fleet design and efficiency optimization~\cite{liang2023ditto, gupta2020architectural, hsia2020cross, gupta2022chasing}. Internally, it can be used for system optimization (e.g., GPU or ASIC accelerator design), performance characterization and analysis, bug reproducibility, etc. It can also be shared with external hardware vendors for early-stage performance testing, evaluation, and joint HW/SW codesign, with minimal infrastructure support and a streamlined IP sharing setup.

The past few years have seen significant advancements in AI benchmarking~\cite{mattson2020mlperf_train, reddi2020mlperf, deepbench, coleman2017dawnbench, adolf2016fathom}. MLPerf, specifically, is an industry-standard suite that covers diverse ML applications, DNN models and optimizers, from training to inference. However, its model diversity and updating speed, as Table~\ref{tab:mlperf_training_benchmarks} shows, cannot match the ever-changing, highly-diverse AI production workloads across cloud infrastructures. Due to workload churn, the execution characteristics of workloads can quickly change over time~\cite{verma2020demystifying}, completely changing the system requirements.


\begin{table}[ht]
    \centering
    \vspace{-0.14in}
    \caption{MLPerf training benchmarks~\cite{mlperf_training_benchmarks}.}
    \resizebox{0.8\columnwidth}{!}{%
    \fontsize{6}{7}\selectfont
    \begin{tabular}{c|c|c}
    \hline\hline
    {\bf Area} & {\bf Model} & {\bf Last updated} \\
    \hline
    Vision & ResNet-50 & May 17, 2021 \\
    Vision & 3D U-Net & Apr 14, 2021 \\
    Vision & Mask R-CNN & Mar 5, 2021\\
    Language & RNN-T & Apr 7, 2021 \\
    Language & BERT-large & May 14, 2021 \\
    Commerce & DLRM & Feb 9, 2021 \\
    Research & Mini Go & Jun 19, 2020 \\
    \hline
    \hline
    \end{tabular}
    }
    \label{tab:mlperf_training_benchmarks}
\end{table}

Additionally, engineers or researchers need to manually select and adapt existing production or open-source workloads to a form that can be used for benchmarking. This process involves a non-trivial investment, since it requires high expertise and comprehensive understanding of the workloads. Also, extracting only the desired components from a production environment can be challenging, since production workloads have many supporting dependencies (e.g., storage, data preprocessing, scheduler), and many proprietary in-house libraries and tooling integration. This can lead to a high cost for maintaining and updating the derived benchmarks to keep up with the fast cadence of AI application design. Therefore, there is a strong need for a new methodology which enables us to efficiently generate AI benchmarks in production scale. 

In this paper, we propose an efficient and scalable framework to create AI benchmarks directly from production workflows in a ``replay as benchmark" manner. We present Mystique, a benchmark generation framework for AI workloads, which leverages the new PyTorch execution trace (ET) capability to record the runtime information of a model at operator granularity, and faithfully replay it to reproduce the original performance. Mystique is efficient and scalable as only a few lines of hook code are needed to collect the traces and generate a benchmark from a production, cloud-scale AI model. Mystique is open-source and publicly available.\footnote{https://github.com/facebookresearch/param.}

Our main contributions are:

\begin{itemize}
\item We build a scalable and automated end-to-end infrastructure that profiles and replays the execution traces from real production AI workloads. 

\item We evaluate Mystique across several production PyTorch workloads running in a warehouse-scale fleet, and show that the generated benchmarks closely match the original, in terms of performance and system-level metrics. 

\item We showcase the portability of the generated benchmarks across platforms and evaluate several use cases Mystique can be applied to, including early stage platform evaluation, subtrace replay, and scaled-down performance testing.


\end{itemize}

\section{Related work}

\subsection{AI benchmarks}

Benchmarks are an easy-to-use representation that captures the most essential characteristics of a workload, while leaving out non-critical aspects of the original application. Benchmarks are powerful tools, as they enable evaluating a target system's performance for a given workload without the need for supporting all dependencies the original application requires, e.g., libraries, upstream/downstream data pipeline setup, job orchestration. These advantages are even more desirable for AI workloads, which are built over complex programming frameworks, have more interconnected SW components, and often run distributedly in cloud environments. 

Therefore, providing a robust methodology to create realistic AI benchmarks that reflect a deployment's behavior is an attractive proposition for all levels of system design, from AI accelerator design to datacenter deployment orchestration.

Unsurprisingly, there have been numerous proposals on benchmarking AI workloads~\cite{mattson2020mlperf, mattson2020mlperf_train, reddi2020mlperf, banbury2021mlperf, deepbench, coleman2017dawnbench, adolf2016fathom, zhu2018benchmarking}. DeepBench~\cite{deepbench} provides a set of basic operations used by deep neural networks (DNN) and evaluates them on different platforms. TBD~\cite{zhu2018benchmarking} identifies eight representative DNN models, and performs a detailed performance analysis on different deep learning frameworks and hardware configurations. DAWNBench~\cite{coleman2017dawnbench} measures the end-to-end performance of training and inference, subject to a specified accuracy, allowing innovation in software, algorithms, communication methods, etc. More recently, MLPerf~\cite{mattson2020mlperf, mattson2020mlperf_train, reddi2020mlperf, banbury2021mlperf} has become an industry-standard suite for ML performance, encompassing a variety of models in different domains (vision, language, recommendation, research) and different deployment scenarios, from datacenter, to edge and mobile. 

While this work has enriched and strengthened the availability of AI benchmarks to the community, their coverage remains limited compared to the vast spectrum of AI workloads deployed in  production. For instance, it is not uncommon to find thousands of AI models in a hyperscaler's fleet at any given time. Curating a small set of benchmarks to approximate general behavior characteristics from such a vast collection is a significant challenge. At the same time, given the rapid pace of innovation in the AI space, new AI workloads appear in datacenters on a daily basis, further hindering the ability of benchmark characteristics to remain up to date. 

In essence, we have a \textit{scalability} issue both in terms of the model space and in terms of time required for benchmark generation. For example, diffusion models~\cite{saharia2022photorealistic, ramesh2022hierarchical} are a new class of generative models that generate diverse high-resolution images, however, have not yet been included in any widely used benchmarks. On the other hand, production models often include adaptations and optimizations on top of open source models customized to their own use cases, and thus can exhibit significantly different performance characteristics from their corresponding open-source versions.

Our insight in dealing with this scalability issue is to rely on automation to generate representative AI benchmarks instead of the current manual curation approach. Mystique enables generating benchmarks at scale with minimal manual input, and provides close behavior resemblance to production flows. 

\subsection{Simulation, emulation, and performance modeling}

Simulation, emulation, and performance modeling offer another way to approximate a workload's performance, when software or hardware is unavailable. Sniper~\cite{carlson2011sniper} is a parallel and scalable CPU simulator using a high-level abstraction for simulating multicore and multiprocessor systems. gem5~\cite{binkert2011gem5} is a modular microarchitectural simulator that has been broadly used for GPU architecture modeling~\cite{beckmann2015amd, power2014gem5, del2006attila}. GPGPU-Sim~\cite{khairy2020accel} provides a cycle-level simulation model of NVIDIA GPUs running CUDA and OpenCL workloads, and enables fast and detailed validation. While these simulation techniques are not specific to AI, they have been extensively used to evaluate software and hardware proposals for AI systems. 

Similarly, there has been a lot of work on performance modeling of ML workloads~\cite{liao2020perfnetrt, yu2021computational, justus2018predicting, li2020characterizing, zhu2020daydream, lin2022building}. Daydream~\cite{zhu2020daydream} predicts model runtime under certain optimizations based on the kernel-task dependency graph. Habitat~\cite{yu2021computational} uses wave scaling and MLPs to predict the execution time of DNN training. Finally, CM-DARE~\cite{li2020characterizing} proposes a performance model for distributed training with cloud-based GPU servers to achieve cost savings and speedup. 

While useful when hardware is unavailable, or requires non-negligible changes, these simulators and performance models still make approximations on the workload behavior and cannot fully capture the complexity of a real system. Also, performance models in particular, usually target specific use cases and cannot easily be extended to a wide range of studies. 

\subsection{Performance cloning and synthetic benchmarks}

Performance cloning is an intuitive way to generate synthetic benchmarks that preserve the performance of real-world workloads. Previous work profiled the architectural characteristics of the target applications, and generated corresponding proxy benchmarks~\cite{ganesan2013automatic, panda2018camp, panda2017proxy, ravi2021micrograd, dangwal2019safer, van2008dispersing, wang2017clone}. MicroGrad~\cite{ravi2021micrograd} collects the CPU metrics and uses a Gradient Descent based tuning mechanism to produce workload clones and stress tests. PerfProx~\cite{panda2017proxy} generates miniature proxies for real-world database applications, based on performance metrics derived from hardware performance counters. CAMP~\cite{panda2018camp} models core performance, memory locality, and the interaction between them to mimic BigData applications. ECHO~\cite{Delimitrou12} focuses on cloning network behavior in distributed cloud applications using statistical models, and generates synthetic benchmarks that resemble the locality and traffic characteristics of the original services. Finally, Ditto~\cite{liang2023ditto} proposes an automated cloning framework that can capture both the low- and high-level performance metrics of distributed cloud applications. 

However, these approaches only focus on CPU performance and miss the critical performance engines exercised by AI workloads, namely GPUs or other accelerators. G-MAP~\cite{panda2017statistical} models the memory access patterns and parallelism of GPU applications to create memory proxies. GPGPU-MiniBench~\cite{yu2015gpgpu} generates miniature proxies of CUDA GPGPU kernels to retain similar performance. However, they focus on only memory or kernel behavior on GPU, and do not consider the sequential execution on the CPU and the interactions between CPUs and GPUs, and therefore cannot reflect the full performance of AI models.



\section{Background}
\label{sec:background}

Finding realistic benchmarks that resemble production cloud workloads is a long-standing problem. Given the limitations of open-source benchmarks and of approaches that rely on simulation or performance modeling, generating synthetic benchmarks that mimic the full stack performance of real applications can enable a wide range of system studies. 

Prior work on performance cloning and benchmark generation mostly focuses on CPU-centric workloads, by collecting their architectural characteristics and generating appropriate assembly instructions. Although, in theory we could apply the same approach to AI applications, this would be overlooking the unique properties that AI workloads exhibit. The fact that most AI workloads are implemented with a handful of frameworks (e.g., PyTorch, Tensorflow, JAX), makes capturing a logically complete yet---representation-wise---succinct and reproducible snapshot possible. In PyTorch, applications invoke low-level operators, such as ATen~\cite{aten}, NCCL~\cite{nccl} to fulfill their execution. By recording the execution information at these invocation boundaries, we can faithfully reconstruct the execution behavior of a complex AI workload. 

We focus on generating benchmarks for PyTorch AI models. We choose PyTorch as our first step because of its widespread use in industry (and our worldwide production environment specifically) as well as academia, and the rich profiling capabilities it offers. Our approach can be extended to support other ML frameworks, as discussed in Section~\ref{sec:support_other_frameworks}. Below we describe what the execution trace (ET) is, and how it enables us to generate realistic AI benchmarks. 

\subsection{Execution Trace (ET)}

\begin{figure}[htb]
    \centering
    \vspace{-0.1in}
    \includegraphics[width=0.34\textwidth]{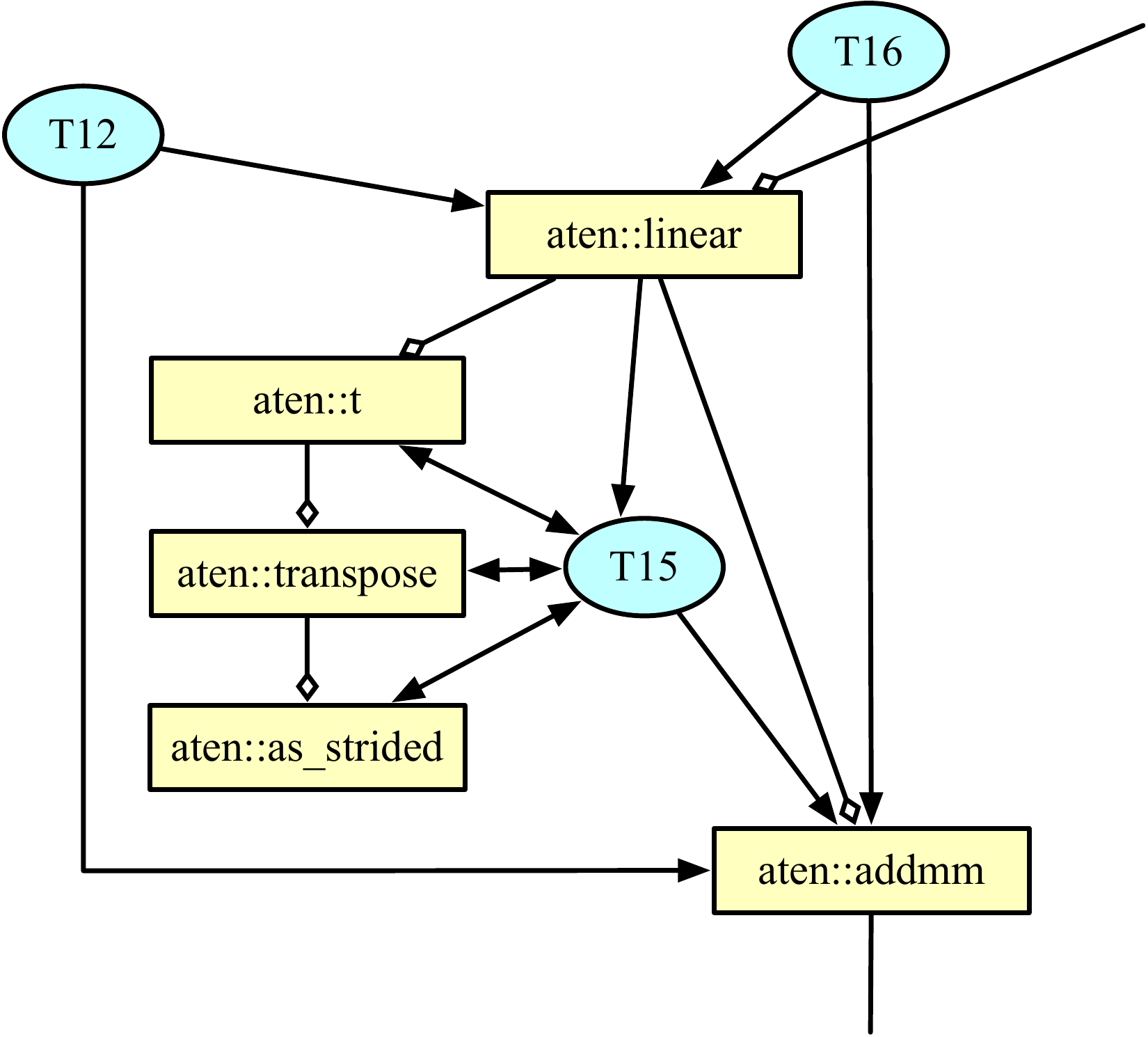}
    \caption{An example of PyTorch's execution trace (the figure only shows a subtrace for simplicity), in which the boxes are PyTorch operators and the ovals are unique tensors. Arrows represent inputs and outputs. Lines ending with a diamond show parent-child relationships between the operators. Execution order is not shown here. }
    \label{fig:execution_graph}
\end{figure}

The execution trace of a PyTorch model is a runtime recording of its operators together with their metadata, such as the execution order, operator schema, input/output arguments, as well as their parent-child relationships. Figure~\ref{fig:execution_graph} shows an example of such an execution trace, where  each node is a PyTorch operator and the connections between the nodes indicate the parent-child relationships, i.e., the calling stack of the operators. Table~\ref{tab:node_schema} shows the key information captured for each node in more detail. 

\begin{table}[ht]
    \centering
    \vspace{-0.1in}
    \caption{Execution trace node schema.}
    \resizebox{0.9\columnwidth}{!}{%
    \fontsize{9}{11}\selectfont
    \begin{tabular}{c|c}
    \hline\hline
    {\bf Key} & {\bf Description} \\
    \hline
    \hline
    \texttt{name} & Name of node \\
    \texttt{id} & Unique ID of this node \\
    \texttt{parent} & Parent node ID \\
    \texttt{op\_schema} & PyTorch operator schema \\
    \texttt{inputs} & Array of input args  \\
    & Actual values for non-tensor args \\
    \texttt{input\_shapes} & Array of input shapes \\
    & Empty [] for non-tensor args \\
    \texttt{input\_types} & Array of input types \\
    & Empty [] for non-tensor args \\
    \texttt{outputs} & Array of output args  \\
    & Actual values for non-tensor args \\
    \texttt{output\_shapes} & Array of output shapes \\
    & Empty [] for non-tensor args \\
    \texttt{output\_types} & Array of output types \\
    & Empty [] for non-tensor args \\
    \hline
    \hline
    \end{tabular}
    }
    \label{tab:node_schema}
\end{table}

Each tensor argument is tagged with a unique ID (a six element tuple) with its shape and data type. This unique ID is used to track the data dependencies among operators and to distinguish each tensor, as we will discuss in Section~\ref{sec:arguments_management}. The execution order across operators is not explicitly recorded but can be inferred from the node IDs, because they are assigned in increasing order, based on execution order. 

\subsection{Advantages of ET}

This execution trace records the metadata that is needed to reproduce the original execution behavior of each operator, offering us a very intuitive way to generate synthetic benchmarks by replaying all operators in the trace according to their original execution order and data dependencies. 

ET stands out among other similar recording approaches because: 1) its API is easy to use and requires minimal source level changes (collection can be enabled by a few lines of code, no performance counters or architectural characteristics needed), 2) its hardware agnostic design makes it portable across different hardware platforms, 3) it incurs very small performance overheads, which facilitates a large-scale automated data collection setup in the background, in a production environment, 4) it has a compactly defined data schema which minimizes the storage support cost in production, and 5) as each ET node is a self-contained entity, the ET format provides great composability which enable more use cases, such as new hardware platform evaluation and scaled-down performance emulation (we discuss this in detail in Section~\ref{sec:use_cases}). These traits make ET ideal for encapsulating broad production workload behavior in an agile manner.


\subsection{PyTorch operators}

Operators are the building blocks in the PyTorch framework, which define the mathematical and logical transformations to be performed on the data. Every operator includes a set of platform-specific implementations, usually written in C/C++ or other domain specific languages, to provide its functionality on the supported hardware. The framework is designed to allow easy custom implementations for reasons spanning from increased performance to enabling new hardware innovations. Among the models we have profiled, we find that operators can be roughly divided into four categories based on the implications each entails when trying to replay them:


\begin{figure}[htb]
    \centering
    \vspace{-0.08in}
    \includegraphics[width=0.44\textwidth]{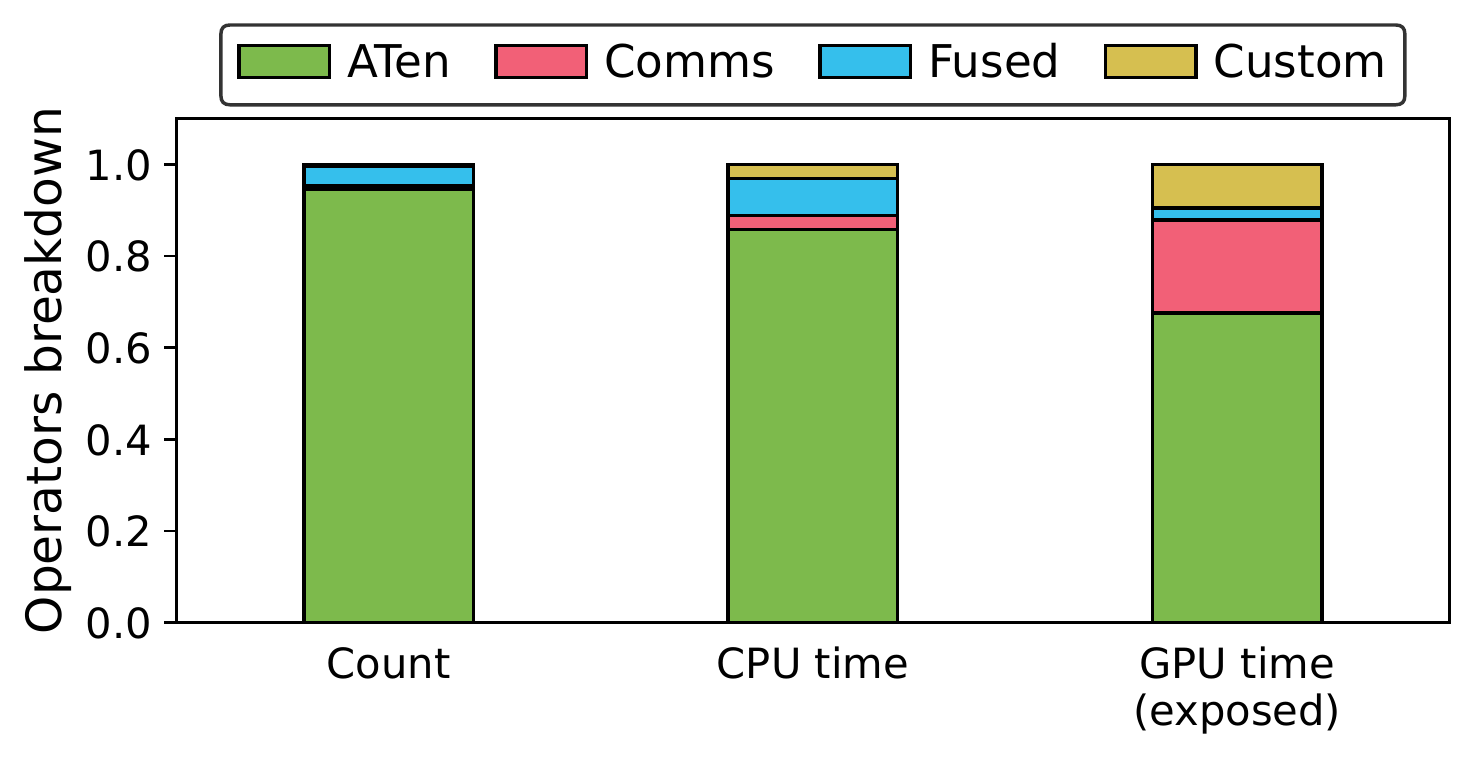}
    \vspace{-0.14in}
    \caption{Fraction of different operators in a production model running on 8 GPUs, in terms of their count, CPU time, and exposed GPU time. }
    \label{fig:operators_breakdown}
\end{figure}

\begin{itemize}

\item \textbf{ATen ops:} ATen is the low-level tensor library and compute backend for PyTorch. It performs the actual computation on tensors, such as addition, matrix multiplication, and batch normalization. 

\item \textbf{Communication ops:} Distributed training across multiple devices has now become the norm to support large scale AI models, as well as to increase the training speed. During distributed training, communication operators are used for synchronization and data transmission among multiple devices. For PyTorch, c10d~\cite{c10d} is the most popular communication library, which offers both collective communication APIs (e.g., \textit{all\_reduce()} and \textit{all\_to\_all())} and P2P communication APIs (e.g., \textit{send()} and \textit{recv()}).

\item \textbf{Fused ops:} Operator fusion is a common optimization technique that merges multiple operators into a single execution instance to reduce the memory access and kernel dispatch overhead. In PyTorch models, it can be easily enabled by applying the \textit{@torch.jit.script} decorator to the model's function definition. There are a couple of available backends in JIT; the default fuser on CPUs is NNC and on GPUs is NVFuser. After PyTorch fuses the original operators, it will emit a single fused operator in place of them during execution.

\item \textbf{Custom ops:} To support the rapidly changing AI landscape, PyTorch provides a user-friendly interface for users to define custom operators. The interface is commonly used to create a novel model building block or to provide a better implementation than the default routines. Using operators imported from other libraries (e.g., FBGEMM~\cite{fbgemm} and torchrec~\cite{torchrec}) in the application is another way to leverage such custom operator support.
\end{itemize}

\begin{figure*}[htb]
    \centering
    \includegraphics[width=0.96\textwidth]{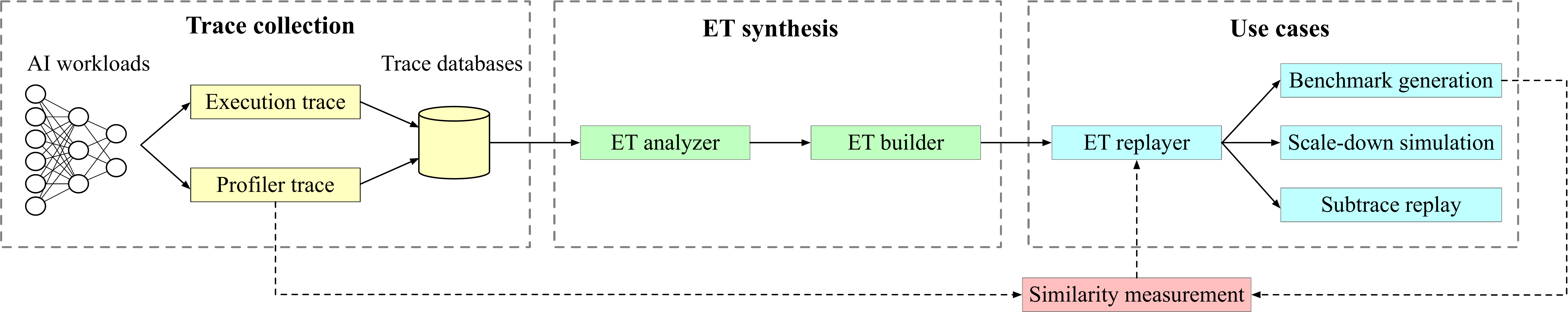}
    \vspace{-0.1in}
    \caption{Overview of our benchmark generation workflow using ET replay. }
    \vspace{-0.08in}
    \label{fig:workflow_overview}
\end{figure*}

Figure~\ref{fig:operators_breakdown} shows the fraction of different types of operators for one of the most popular production workloads running in our warehouse fleet. We show the ratios across three metrics: operator count, CPU time, and GPU time, which correspond, respectively, to the number of occurrences of the operators, the execution time on CPUs, and the execution time of the kernels launched on GPUs. In particular, for communication operators, we measure the exposed GPU time, which is the time that their launched kernels are not running in parallel with any other computation kernels. As the default compute backend of PyTorch, ATen operators take up the lion share in terms of all three metrics. Fused operators are the second in count but have the shortest GPU time. Custom and communication operators are quantitatively modest, but have long GPU time; the former are usually complex in functionality and therefore expensive to execute, and the latter can also come at a significant cost in large-scale distributed deployments. 

Considering the high fraction of ATen and communication operators, we mostly focus on them during the operator reconstruction phase, discussed in the next section. 

\section{Mystique Design}
\label{sec:eg_replay_as_benchmarks}


Figure~\ref{fig:workflow_overview} shows a high-level overview of Mystique, our benchmark generation framework based on ET replay. First, we collect both the execution trace and the profiler trace of the fleet's AI workloads under live traffic. Then the ET analyzer and builder preprocess the traces and select the most commonly-occurring ones. Currently, we pass these ETs to the replayer in their original form and in the future we plan to add more sophistication to accommodate additional uses, such as operator obfuscation for enhanced IP protection. Finally, the ET replayer gets the input traces and creates the desired benchmarks or configures them for different use cases. The whole workflow is fully automated, so we can constantly update the benchmarks using the latest collected traces without any human involvement. Additionally, we add the feedback loop between the replay and original traces by comparing their similarity to validate and improve our methodology. In the rest of this section, we describe the trace collection, and each essential part of the replay-based benchmarking methodology. 

\subsection{Trace collection}

To collect the ET of a PyTorch model, a user currently needs to insert simple hooks into the code to instantiate an \textit{ExecutionGraphObserver} and use \textit{start()} and \textit{stop()} methods to control when the execution is recorded. Typically, we only need to collect a single execution iteration, requiring tens to hundreds of milliseconds, since the execution trace of a model is mostly the same across iterations. Each process has a single observer instance and if running under a distributed setup, multiple execution traces will be collected, one for each process. These traces need to be collected from the same iteration, to ensure that the same communication operators are recorded. The alternative can lead to deadlocks during replay, if we cannot match the communication operators across ETs. 

\begin{python}
from torch.profiler import ExecutionGraphObserver

# Instantiate the runtime observer
et_path = "/tmp/execution_trace.json"
et_observer = ExecutionGraphObserver()
et_observer.register_callback(et_path)

# Insert hooks into execution (e.g., training) loop
def training_loop():
  # Collects profiler trace
  with torch.profiler.profile(
    activities=[ProfilerActivity.CPU, 
                ProfilerActivity.CUDA],
    on_trace_ready=profiler_trace_handler,
  ) as pf:
    for idx in range(100):
      if idx == 10:
        # Start ET capture
        et_observer.start()
      if idx == 11:
        # Stop ET capture
        et_observer.stop()
      model.step()
      pf.step()
\end{python}

In some cases, collecting the ET alone is not enough to fully reproduce the behavior of a workload, as it lacks the CUDA stream execution information. In those cases, we combine ET with another runtime trace, collected by the PyTorch Profiler\cite{pytorchprofiler}. We discuss this in Section~\ref{sec:parallel_streams}. 


The pseudocode of trace collection is shown above. Adding the ET observer and profiler to record the runtime information can introduce some overheads to the performance of the original AI model. However, this overhead is small and only occurs once, and it does not affect the accuracy of the generated benchmark, as our replay method does not rely on any temporal information captured in these traces or hardware-level performance metrics. 

\subsection{Operators selection}

Given an execution trace, we need to select which operators to replay, because some of them are redundant. For example, \textit{aten::linear()} has included two of its child operators \textit{aten::t()} and \textit{aten::addmm()} as part of its implementation. At runtime all three operators will be caught in the ET, however, we only need to replay the parent one, which in this case is \textit{aten::linear()}. To identify these redundant operators, since the parent operator is always executed before its children, we can traverse the operators in the order of execution, keep each operator we encounter and skip all its child operators, based on the parent-child relationships captured in the execution trace. 

\subsection{Operators reconstruction}





\subsubsection{ATen operators}

We reconstruct each ATen operator through the TorchScript IR (Intermediate Representation) using its captured operator schema in ET, which includes the operator name and data types for its input and output arguments. We implement a string-based parser to extract this key information from the schema, which is then used to build the canonical textual representation of the IR. Finally, we compile the IR with TorchScript to create the callable function for each operator to use during replay. The pseudocode below demonstrates this procedure: 

\begin{python}
# Captured op schema in ET
op_schema = "aten::add.Tensor(Tensor self, 
             Tensor other, *, 
             Scalar alpha=1) -> Tensor"

# Extract name and arguments from schema 
op_name, op_args = parser(op_schema)

# Build IR with extracted information
torchscript_ir_str = builder(op_name, op_args) = ""
    graph(
    return (
""

# Compile IR to callable function
graph = torch._C.parse_ir(torchscript_ir_str)
cu = torch._C.CompilationUnit()
func = cu.create_function(op_name, graph)
\end{python}



\subsubsection{Communication operators}

To reproduce the communication pattern of the original workload, we need to replay the communication operators with their original arguments, such as the process group and message size. The metadata can be obtained from the execution trace; for the execution of the operators during replay, we leverage the existing PyTorch distributed infrastructure and implement a wrapper over the low-level interfaces to pass the appropriate parameters. We create new process groups and map them to the original groups, and for each operator, we select the same data type and size as the original, to ensure a similar communication pattern in replay. Depending on the operator's execution mode, we wait for it to complete if it is blocking, or execute it asynchronously by registering its callback to check back later. 

\subsubsection{Custom operators}
\label{sec:custom_ops_reconstruction}

Custom extension is a mechanism that PyTorch developed to allow users to create their out-of-source operators, distinct from the default backend. Similar to the other operators, their input and output arguments are captured in the execution trace, but that is not sufficient for replay, as we do not know their specific implementations. To handle this case, we expose an interface, which allows users to register their custom operators together with their implementations. During replay, we look up the registry and use the provided implementation to replay each custom operator.

\subsubsection{Fused operators}

Pointwise operators can be fused into a single kernel to amortize memory access time and kernel launch time. However, the current implementation of execution trace does not support the metadata of fused operators. For now, we skip these operators altogether because they comprise only a small percentage of the operator list, and have negligible impact to performance, as shown in Fig.~\ref{fig:operators_breakdown}. Fusion behavior can be reproduced via PyTorch JIT and we plan to add this support once its reproduction metadata is available in ET.

All operator reconstruction happens during the initialization phase of the replay, such that they can be directly invoked in the real execution to avoid any runtime overhead. 

\subsection{Arguments and tensor management}
\label{sec:arguments_management}

In addition to the functionality, input arguments also play an important role to the performance of an operator. For arguments with basic types, such as \textit{int} or \textit{bool}, we can simply save their values and reuse them during replay, however, for part of the tensors we need to instantiate them in advance. 

If we track the occurrence of all tensors that are used as input, we can divide them into two categories. We call one type \textbf{intermediate tensors}, which are generated as the output of an operator executed earlier, before being used. The other category are \textbf{external tensors}, whose generation is not observed in the execution trace. This classification can be done by tracking the appearance of each tensor based on its unique tensor ID, during the trace traversal in execution order. For \textbf{intermediate tensors}, we save them at the time of generation, and pass them to the downstream operators, according to the data dependencies between operators. For \textbf{external tensors}, we explicitly instantiate them before execution. 

By default, we instantiate a tensor with the same shape and data type as the original but with random values, as the performance of an operator is not related to the values of the input tensors. We find that this holds true for most operators, as we will show later in the evaluation section, minus a few rare exceptions. One case we have met is the \texttt{lookup operator} for embedding tables. One of its input tensors stores the lookup indices, whose value directly determines the access pattern and has a strong correlation with performance. In this case, we would need to specify the values for that tensor based on some additional information, such as the table size, indices distribution, or pooling factor. Since not all this information is captured in the ET, for now, we set the default values for the missing information empirically, derived by the operators in our production environment, and we additionally provide an interface for users to further refine them. We leave automatically processing such special cases to future work. 

\subsection{Parallel stream execution}
\label{sec:parallel_streams}


A CUDA stream is a sequence of kernels that execute in issue-order on the GPU, and CUDA applications are allowed to launch kernels concurrently on different streams to improve device utilization and execution efficiency. The most representative scenario is the parallel execution of computation and communication kernels to hide the networking overhead. Also, data transfer between the host and device is often optimized to run on a separate stream. A model's stream execution pattern can have a significant impact on its performance, and we need to take this into account during the benchmark generation.


To do so, we need to identify which stream each kernel is executed on, and which operator launches each kernel, so that we can prepare multiple streams and dispatch each operator to its corresponding stream during replay. Unfortunately, current ET does not include any stream or kernel information so we need to extract this from another runtime trace, collected via the PyTorch profiler\cite{pytorchprofiler}. This profiler has been broadly used in the PyTorch community and the extraction can be easily performed based on the launching relationships between the operators and kernels. For now we use this trace as an ET enhancement, and based on our feedback, the ET working group is actively working on integrating the kernel information into the ET representation. 

\subsection{Putting it all together}

Our ET replay approach first collects the execution trace and profiler trace of a model to capture both the operators with their metadata, and their launched GPU kernels. We then walk through the trace according to the execution order, distinguish individual tensors, and identify the operators to replay. Next, we reconstruct the callable function for each operator, prepare the necessary tensors, and initialize the distributed environment, if necessary. Finally, we replay the operators on different streams with the same execution order, input arguments (but not values for tensors), and data dependencies as in the original workload, to faithfully reproduce their original performance characteristics. In case of a distributed deployment, for validation purposes, the same number of processes will be spawned as the original, each using a separate execution trace, and repeating all the steps above. We also enable scaled-down replays of an AI workload without the need for retraining, as discussed in Section~\ref{sec:scale_down_emulation}.

\section{Implementation}

Mystique is built on PyTorch in approximately 8,000 LoC. It currently supports all basic ATen operators, a large fraction of custom operators used in our production workloads, a few common libraries like FBGEMM, and the c10d distributed library with all four types of backends (nccl, gloo, mpi, ucc). 

To leverage Mystique, it is required that users have access to the source code to insert hooks into their PyTorch model to collect the execution and profiler traces; typically around 10 lines of added code. Although the profiling process incurs some runtime overhead, its impact is negligible, since we only need to trace a single iteration. We have also broadly tested it in our production deployment at Meta, and have not observed any noticeable reliability change. Our framework then takes the traces as input, follows the steps we discussed in Section~\ref{sec:eg_replay_as_benchmarks} to analyze the traces, and generates a single PyTorch program as a benchmark. The program contains operators with a hardcoded execution order, input arguments and data dependencies, and can directly run on any platform as a normal PyTorch application. For the distributed training deployment, we use mpirun~\cite{mpirun} to create the distributed environment and spawn multiple processes, each of which uses its own input traces to generate and execute the benchmark. Synchronization and data sharing between processes is automatically achieved by the communication operators.

Given the granularity and flexibility of the execution trace, our ET replay method can be used to explore more use case scenarios, in addition to completely replicating the original behavior, as we discuss in Section~\ref{sec:use_cases}.

\section{Evaluation}

\subsection{Platforms}

We evaluate Mystique on a production cluster of 10 servers with NVIDIA Tesla A100 GPUs and V100 GPUs, and unless specified, the results are collected on A100 GPUs. We use CUDA 11.4 and PyTorch 1.14 as our testing environment. 

\subsection{Workloads}
We focus on the following four popular and representative models in their respective fields, but have also validated Mystique against many other AI models, with similar results.

\begin{itemize}

\item \textbf{PARAM linear}: PARAM~\cite{param} is a benchmark suite of compute and communication microbenchmarks, as well as full workloads for both training and inference. We select a representative linear model with 20 linear layers and set batch size to 512 and data type to float32.

\item \textbf{ResNet:} We choose the ResNet18 model from torchvision~\cite{torchvision}, with batch size 128 and data type float32. For its distributed deployment, we use the default Distributed Data-Parallel (DDP) training framework by PyTorch.

\item \textbf{ASR:} We use a production multi-GPU automatic speech recognition (ASR) training flow implemented with the Fairseq~\cite{ott2019fairseq} toolkit. At its core, ASR is a neural-network-based acoustic model. 

\item \textbf{RM:} RM is a leading edge multi-node, multi-GPU production recommendation model that pushes the boundary for large-scale training. It is the production implementation that the open-source DLRM benchmark~\cite{naumov2019deep} aims to approximate. In our experiments, several configurations of this model are used to cover different workload setups (e.g., different distributed training set sizes). 

\end{itemize}

\subsection{Operators coverage}

\begin{table}[ht]
\vspace{-0.1in}
    \centering
    \caption{Ops coverage rate across evaluated workloads.}
    \resizebox{0.95\columnwidth}{!}{%
    \fontsize{5}{8}\selectfont
    \begin{tabular}{c|c|c}
    \hline\hline
    \multirow{2}{*}{\bf Model} & \multicolumn{2}{c}{\bf Operators coverage} \\
    \cline{2-3}
    & {\bf Count} & {\bf Execution time} \\
    \hline
    PARAM linear & 100\%   & 100\%  \\
    ResNet       & 100\%   & 100\%  \\
    ASR     & 99.6\%  & 75.7\% \\
    RM         & 96.8\%  & 90.9\%  \\
    \hline
    \hline
    \end{tabular}
    }
    \label{tab:ops_coverage}
\end{table}

\begin{figure*}[tb]
    \centering
    \includegraphics[width=0.9\textwidth]{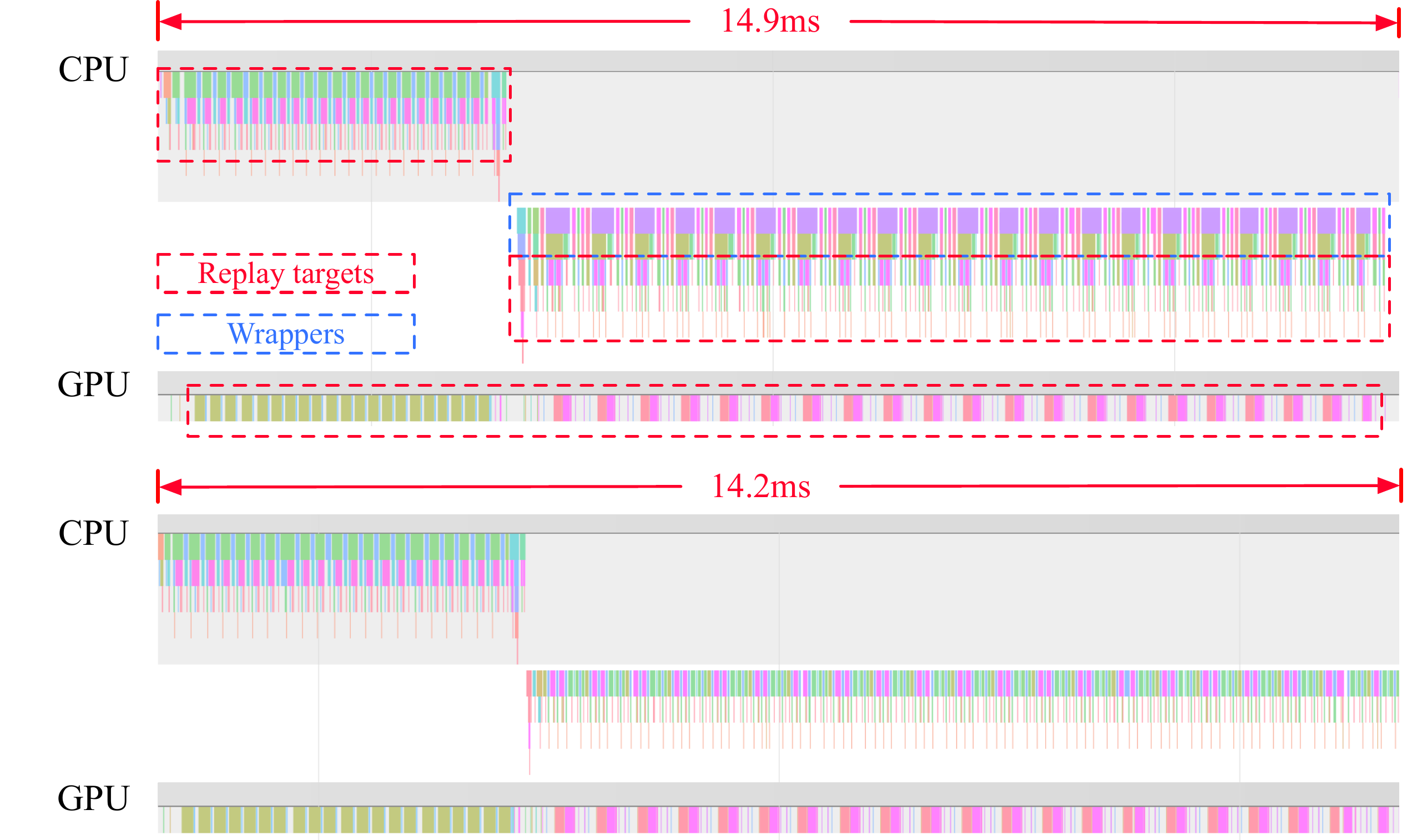}
     \caption{Runtime profiler traces of PARAM linear (top) and its benchmark (bottom) for a single training iteration. Within each trace, the bars on the top are PyTorch operators executed on CPU and the bars at the bottom are GPU kernels, with length indicating the real execution time. The color of bars represents the operator type and color matching shows type matching. }
    \label{fig:resnet_trace}
\end{figure*}

Table~\ref{tab:ops_coverage} shows the current operator coverage rate for our framework across the different studied workloads. The coverage rate denotes the percentage of operators that we are able to replay compared to the total number of operators in a workload. We show the fractions in terms of both count and execution time. Since we support all ATen operators, which are the compute backend of PyTorch, we can achieve a very high coverage rate on the operator count for all workloads. Two of our production workloads have a relatively lower coverage in terms of execution time, since we are currently missing support for fused operators and a subset of custom operators, with the latter dominating the execution time gap. These custom operators normally perform very specific tasks, for example, a LSTM network in NLP models. We provide users with a programmable interface to register more of their custom operators, which can lead to higher coverage and accuracy. 

\subsection{End-to-end execution time}


Figure~\ref{fig:resnet_trace} shows the runtime traces of a single training iteration for the PARAM linear model (top), and its replayed benchmark (bottom). Within each trace, we separate execution time between threads running on the CPUs (top block for each trace) and kernels running on the GPUs (bottom block).

In the original workload's trace (top), there are two threads on the CPUs since the backward operators are automatically performed by PyTorch's autograd engine on the other thread, and we similarly use two threads in replay. The overall execution time of the operator sequence in the replayed benchmark is 14.2 ms, very close to the original's 14.9 ms. When zoomed in, we can see that the execution time of each individual replayed operator, i.e., the length of each bar, and the execution pattern across operators, i.e., how bars interleave with each other, is very similar to the original. The vertical height of the bars is determined by the operator's call stack depth. The small difference in height between original and replayed is due to additional wrappers like autograd::engine::evaluate\_function in the original model, which do not perform any meaningful work; in the synthetic model we only replay their underlying operators (``Replay targets'' regions). 


\begin{table}[ht]
    \centering
    \caption{E2e execution time of a single iteration. }
    \resizebox{\columnwidth}{!}{%
    \fontsize{9}{12}\selectfont
    \begin{tabular}{cccc}
    \hline\hline
    {\bf Model} & {\bf Original} & {\bf Original}  & {\bf Replay} \\
          &          & (exclude unsupported) & \\
    \hline
    PARAM linear & 14.9ms   & 14.9ms & 14.1ms  \\
    ResNet       & 64.4ms   & 64.4ms & 70.7ms  \\
    ASR     & 316.3ms  & 239.3ms  & 229.1ms \\
    RM          & 65.9ms   & 59.9ms  & 58.4ms  \\
    \hline
    \hline
    \vspace{-0.3in}
    \end{tabular}
    }
    \label{tab:e2e_exeuction_time}
\end{table}

\begin{figure*}[htb]
    \centering
    \vspace{-0.18in}
    \includegraphics[width=0.96\textwidth, height=3.7cm]{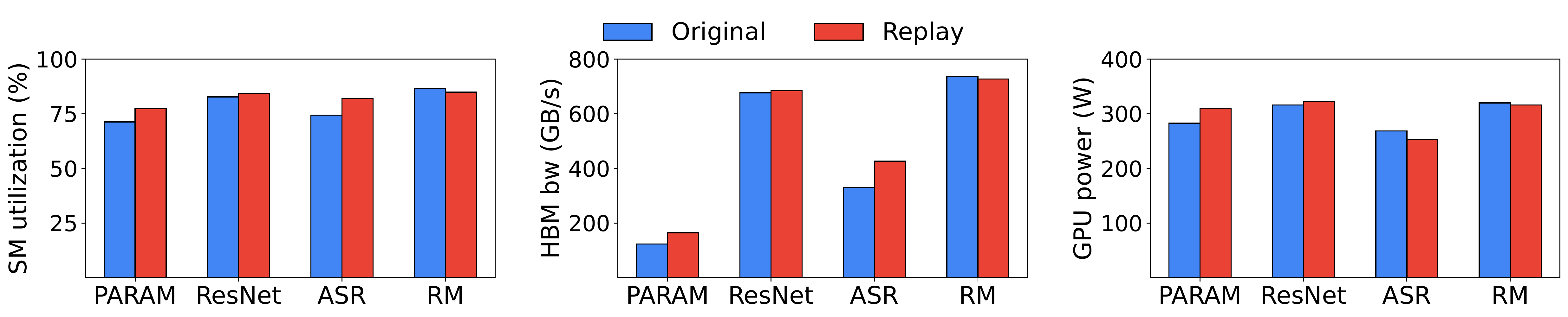}
    \vspace{-0.12in}
     \caption{Comparison of SM utilization, HBM bandwidth, and GPU power for each model and their replay counterparts. }
    \vspace{-0.08in}
    \label{fig:system_metrics}
\end{figure*}

Table~\ref{tab:e2e_exeuction_time} shows the original and replayed execution time for a single iteration of each workload running on a single GPU. For a fair comparison, we also include the original execution time that excludes the unsupported operators, and we use this calibrated execution time for the rest of our evaluation. This table shows to what extent Mystique captures the covered operators' execution time. We obtain high accuracy across all applications, with 5.4\% error on PARAM linear, 9.8\% error on ResNet, 4.3\% error on ASR and 2.5\% error on RM, when comparing the overall performance of all replayed operators. 




\subsection{System-level metrics}

In addition to execution time, system-level metrics are also important for a benchmark generation framework to capture, to ensure high similarity with the original workload. Specifically, for these AI workloads, we are more interested in GPU-related metrics, including both macro- and micro-level characteristics.

\subsubsection{Macro-level}

Figure~\ref{fig:system_metrics} shows three representative and widely evaluated metrics for AI workloads, in production environments: SM utilization, HBM bandwidth, and GPU power, across all workloads and their replay counterparts. All models are using a single A100 GPU. We collect these metrics over thousands of iterations, and show their average values. Compared to the other three workloads, RM has the highest resource utilization, and therefore the highest power usage. The HBM bandwidth gap of ASR is a little larger than the others, due to the small number of custom operators we do not yet support. The results illustrate that different workloads exhibit very different performance characteristics at a macro level, but can be accurately captured by our replay methodology, and reproduced in the generated benchmarks.

\subsubsection{Micro-level}

\begin{figure}[htb]
    \centering
    \includegraphics[width=0.485\textwidth]{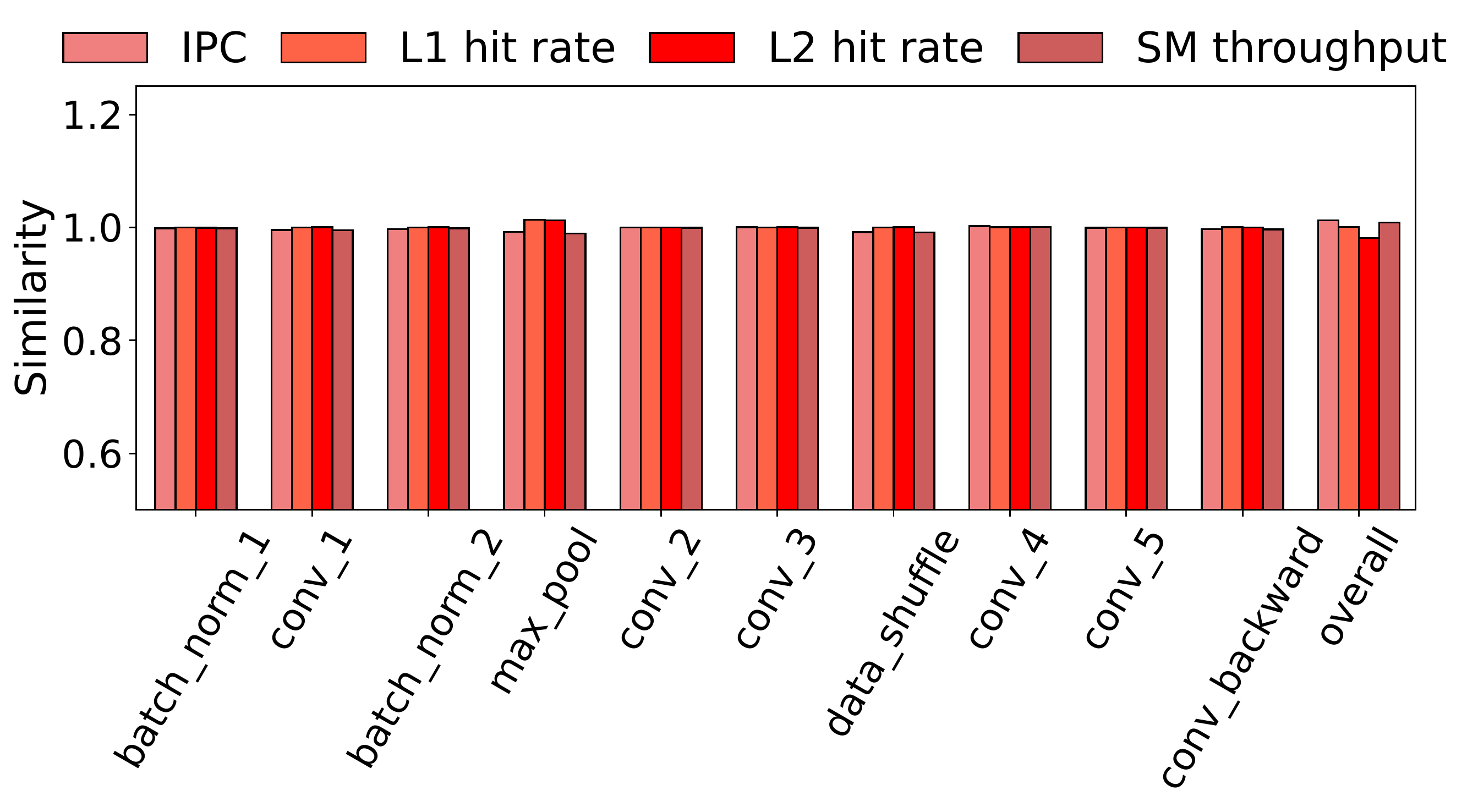}
    \vspace{-0.24in}
     \caption{Normalized per-kernel and overall micro-level architectural metrics. }
    \label{fig:low_level_metrics}
\end{figure}

In Figure~\ref{fig:low_level_metrics}, we show the similarity between the original ResNet and the generated benchmark, in terms of its fine-grained performance characteristics, including IPC, L1 hit rate, L2 hit rate and SM throughput. We choose the top 10 CUDA kernels in terms of runtime and the overall performance among all kernels, and normalize the data to that of the original model. The top 10 kernels account for 50.3\% of the total execution time, and the overall deviation across all kernels is all within 2\%. The results demonstrate that Mystique can also faithfully clone microarchitectural characteristics, which is inherited from our accurate replication of the model at operator level.


\subsection{Distributed training}

Distributed training is now very common practice for AI workloads, as their model sizes and datasets keep rapidly growing. To evaluate the scalability of our ET replay-based framework, we collect the runtime traces of the RM workload running on 8 nodes with 64 NVIDIA A100 GPUs total, and interconnected via NVLink (intra-node) and a 200Gbps NIC per GPU (inter-node), and then replay it under the same setting. To enable large-scale execution, we adjust RM's parameters, resulting in variations in behavior when compared to the single-GPU version.

Table~\ref{tab:multiple_nodes} shows the execution time per iteration and the system-level metrics per GPU, averaged across the profiling duration and all 64 GPUs, for the original model and the replayed benchmark. The performance of the generated benchmark is very similar to that of the original, with a slight difference primarily attributed to the inaccuracy of a few communication operators replay. This evaluation validates the scalability of Mystique for large-scale distributed deployments.


\begin{table}[ht]
    \centering
    \caption{Scalability evaluation on 8 nodes with 64 GPUs.}
    \resizebox{0.95\columnwidth}{!}{%
    \fontsize{5}{8}\selectfont
    \begin{tabular}{c|c|c}
    \hline\hline
    {\bf Metric} & {\bf Original} & {\bf Replay} \\
    \hline
    Execution time (ms) & 102.5  & 113.1  \\
    SM utilization (\%)       & 49.6   & 43.6  \\
    HBM bandwidth (GB/s)     & 418.5  & 364.3 \\
    GPU power (W)         & 228.1 & 204.8 \\
    \hline
    \hline
    \end{tabular}
    }
    \label{tab:multiple_nodes}
\end{table}


\subsection{Cross platform validation}

Mystique operates at operator-level to reproduce the performance and resource characteristics of an original AI workload. This hardware-agnostic operation allows the generated benchmark to be portable across platforms without regeneration. To validate this, we test the performance of all four studied workloads and their corresponding generated benchmarks on three platforms: Intel Xeon Platinum CPU, NVIDIA Tesla V100, and NVIDIA Tesla A100. We only use the trace collected on the A100 server to generate the synthetic benchmarks, and then run them across the different platforms.

\begin{figure}[htb]
    \centering
    \vspace{-0.14in}
    \includegraphics[width=0.495\textwidth, height=6.5cm]{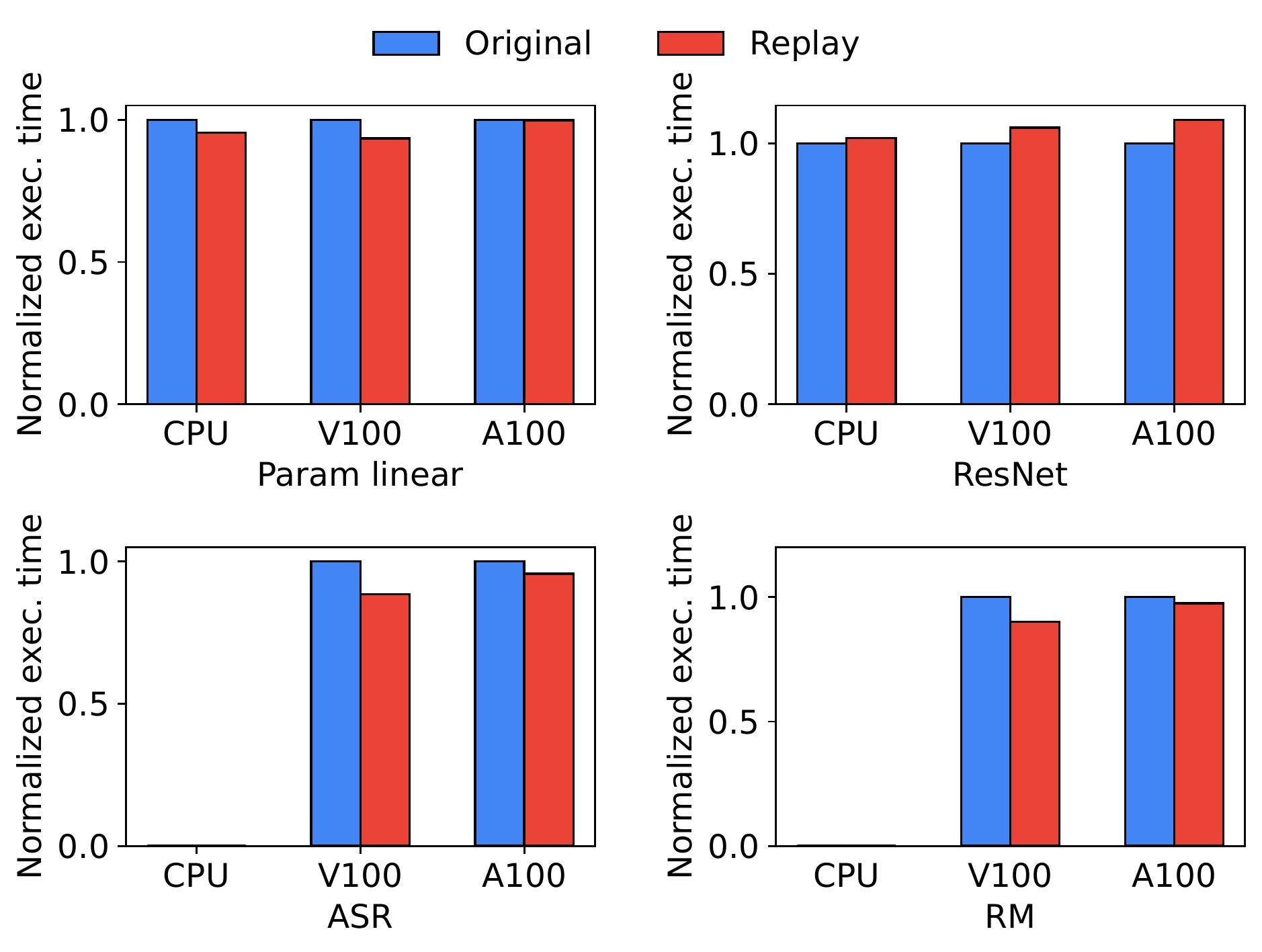}
    \vspace{-0.2in}
     \caption{Normalized execution time of all workloads and their replayed benchmarks on different platforms. }
    \label{fig:cross_platform}
\end{figure}



Figure~\ref{fig:cross_platform} shows the validation results of all four workloads. We normalize the execution time of the replayed benchmark to that of the original workload on each platform. For two production workloads ASR and RM, we only show their performance on the two GPU platforms, as they cannot directly run on CPU. The figure shows that execution time between original and replayed application matches across platforms, demonstrating the portability of our replay methodology. 

\subsection{Power efficiency sensitivity sweep}


\begin{figure}[htb]
    \centering
    \includegraphics[width=0.489\textwidth, height=5.7cm]{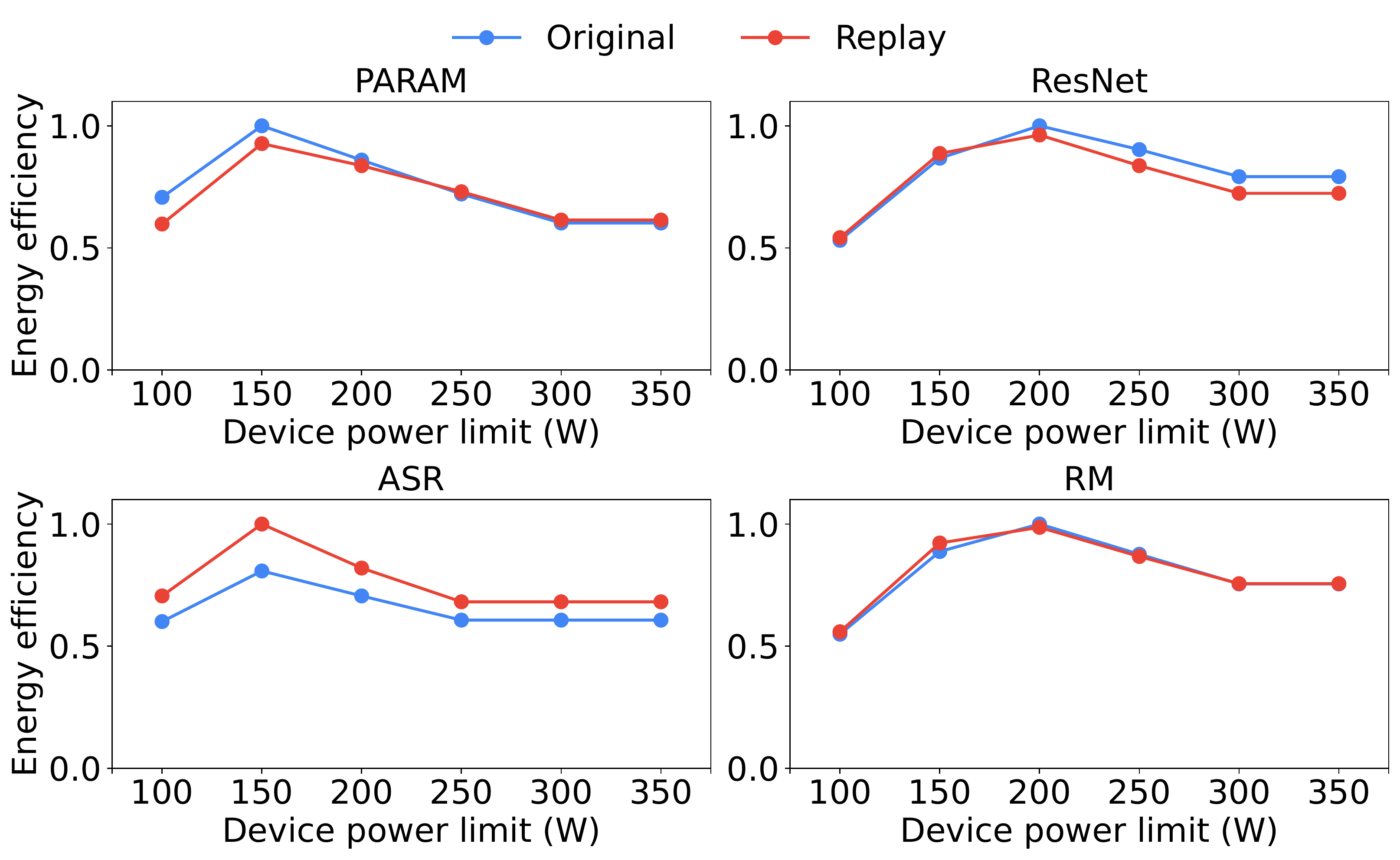}
     \caption{Normalized energy efficiency under varying power limit. }
    \label{fig:energy_efficiency}
    \vspace{-0.14in}
\end{figure}

Power efficiency is an important metric in many system designs, and large-scale training is no exception. Due to the sheer size of our production training fleet, even a small power efficiency gain translates to huge infrastructure cost savings. Here we demonstrate that the benchmark generated by Mystique is able to mimic the power efficiency characteristics of the original application, when we sweep certain system design knobs (the device power limit in this example). 

Figure~\ref{fig:energy_efficiency} displays the power efficiency sensitivity curves of all studied workloads and their corresponding synthetic benchmarks, as we set the GPU power limit to different levels. The x-axis is the GPU power limit we want to sweep, and the y-axis is the normalized power efficiency, defined as the throughput over power. Our generated benchmarks closely track the sensitivity trend of the original workloads, demonstrating that such a methodology can be effectively used to evaluate system performance in the place of real workloads, when they are not available. 


\section{Use Cases}
\label{sec:use_cases}

By leveraging the execution trace (ET), Mystique opens up many opportunities on how to conduct AI system evaluation. In this section we describe several use cases we experimented with using our current framework. 

\subsection{Subtrace replay}

The execution trace is made up of nodes (operators) connected through parent-child relationships; this composability allows us to replay only a subtrace of interest or certain types of operators, enabling fast and efficient testing of a specific component instead of the entire model.

\begin{figure}[htb]
    \centering
    \includegraphics[width=0.48\textwidth, height=2.8cm]{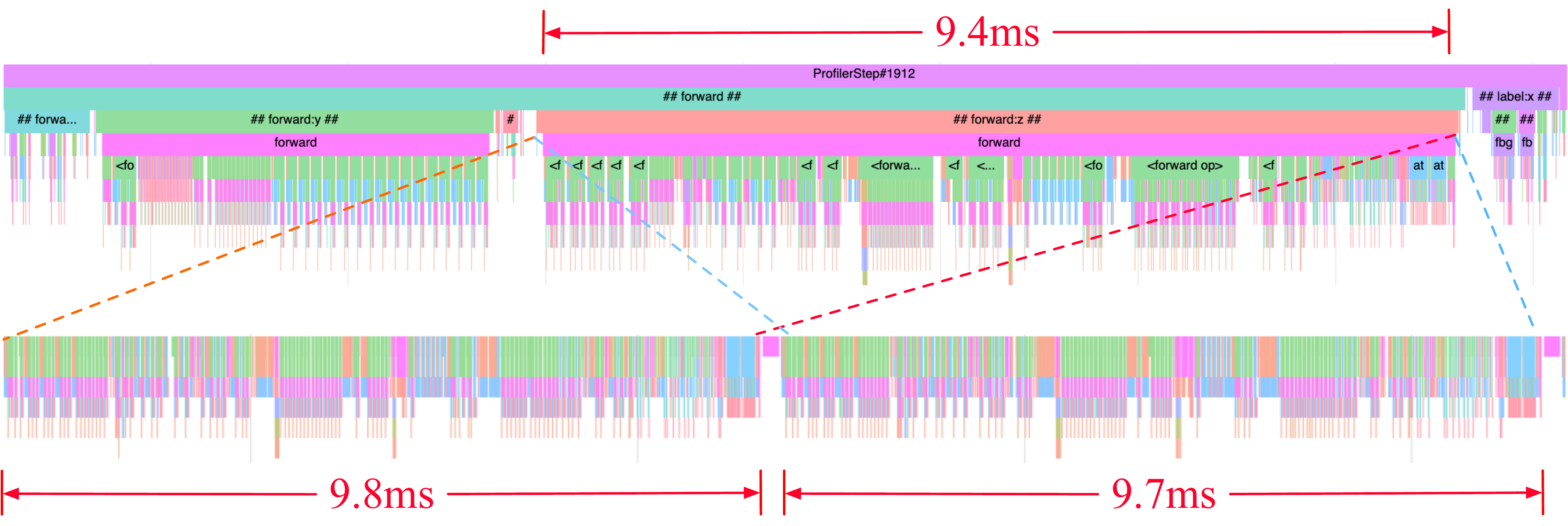}
    \caption{Runtime traces of original model (top) and two iterations of the subtrace replay (bottom). The original label names are replaced for confidentiality reasons. }
    \label{fig:subgraph_replay}
    \vspace{-0.14in}
\end{figure}

To do this, a user can leverage the \textit{record\_function} context manager~\cite{pytorchprofiler} in the PyTorch profiler to label an arbitrary range of code with a user-defined name. Then in the ET, a new operator will appear as the parent of all operators within that code range. When traversing the trace to select which operators to replay, we can use this name to easily locate this operator, and only replay the subtrace underneath. In Figure~\ref{fig:subgraph_replay}, the subtrace located under the operator \textit{\#\# forward:z \#\#} is selectively replayed for the RM workload to measure the performance of this specific segment of forward process. The repeated replay traces in the bottom verify that we are only replaying the target subtrace, and the results demonstrate that the original performance of this subtrace is captured.

Similarly, by filtering operators based on their types, our framework can also be easily configured to replay only specific types of operators. For example, we have used it to quickly examine and locate network issues in our production environment, by replaying the communication operators exclusively.

\subsection{Early stage platform evaluation}

Benchmarks are an essential tool to evaluate the performance of new hardware platforms, especially when the full software environment is not yet fully equipped. While simple microbenchmarks can provide some indication of performance, using a production-like benchmark that exercises the full stack in a similar way to the original workload lends significantly more confidence to the results. However, in the early stages of a new platform design, it is often difficult to run an exact copy of a complex AI model on it. Manually forcing the full application stack to run can be time-consuming, cumbersome, and error-prone.

\begin{figure}[htb]
    \centering
    \includegraphics[width=0.4\textwidth, height=4cm]{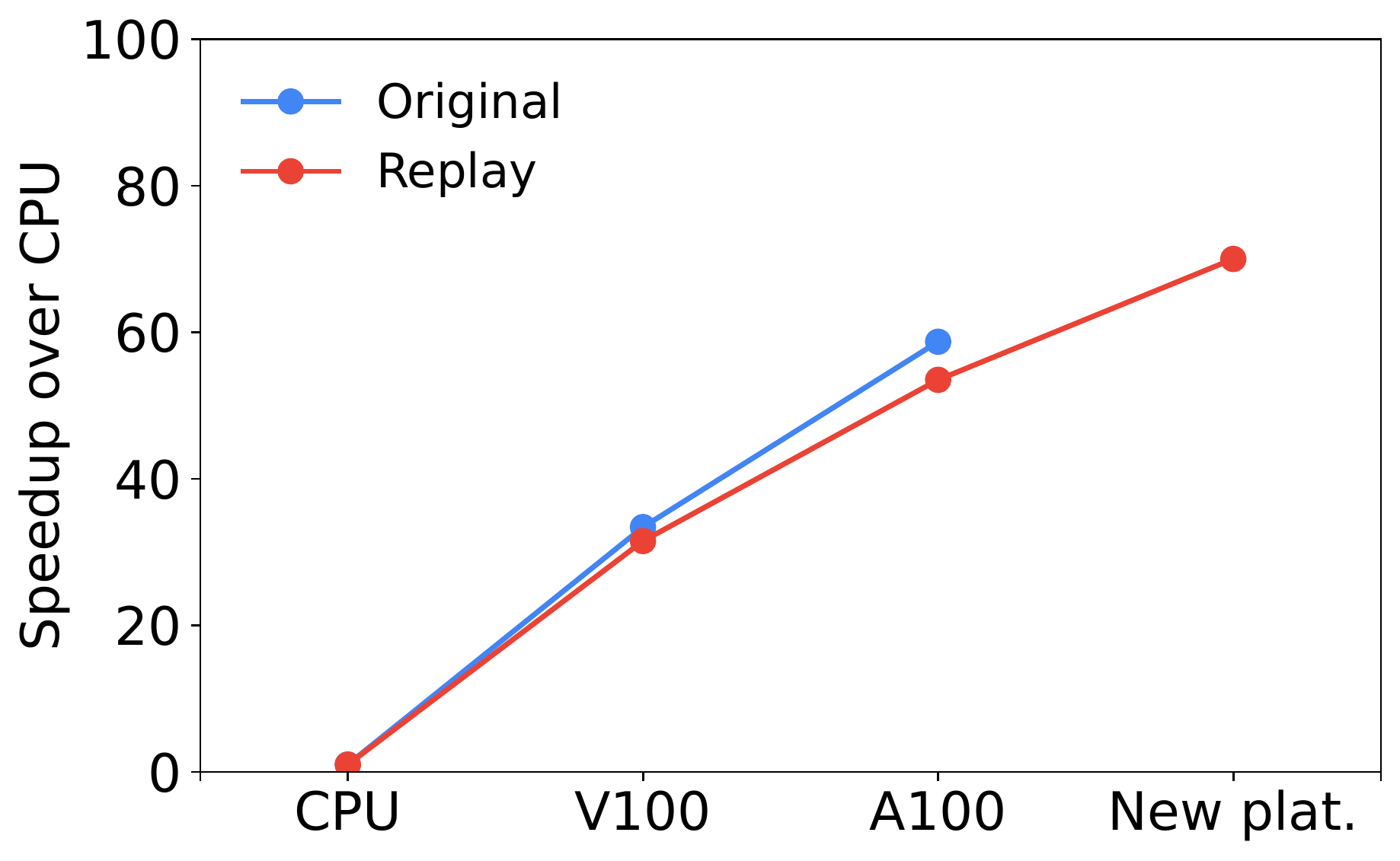}
    \vspace{-0.14in}
    \caption{Execution time speedup for new, experimental platform over CPU. }
    \vspace{-0.14in}
    \label{fig:early_stage_evaluation}
\end{figure}

Fortunately, Mystique provides a solution to this problem, as it has minimal software dependencies and can be easily modified at the operator level to skip unsupported operators. As illustrated in Figure~\ref{fig:early_stage_evaluation}, when evaluating a new platform with limited software installed, such as only the OS and necessary libraries like PyTorch and CUDA, but without many in-house dependencies, libraries and tools, our synthetic benchmarks can be used to accurately infer the potential performance benefit of the new platform, as indicated by the red line.


\subsection{Scaled-down performance emulation}
\label{sec:scale_down_emulation}

DL models have been growing exponentially in terms of their size, complexity, and data requirements, leading to the rapid adoption of large-scale training deployments in production environments. Using models that require hundreds of GPUs for training is now common in AI use cases, but evaluating their performance is challenging due to the high cost and complexity of deploying and supporting large-scale testing setups. Therefore, there is a need to evaluate a model's large-scale training performance using a much smaller setup.

Distributed data-parallel training involves workers performing the same computation on their assigned data chunks, with the local computation remaining unchanged with the number of workers. The network communication, however, is the main performance factor that varies with scale. Benchmarking at the granularity of operators allows for evaluation of large-scale executions using a smaller scale, achieved by adjusting the communication cost during replay. This requires no changes to the original training implementation, avoiding the need for domain specific knowledge and coding changes.

Our initial approach involved adding dummy delay to the communication path to account for the mismatch between small-scale testing and large-scale deployment. The delay was determined empirically based on the network cost model. We demonstrated the feasibility of this approach by successfully reproducing the execution time of the 64 GPUs RM model training using only 2 GPUs. Although there are opportunities to improve upon this simple approach, exploring them fully requires significant engineering effort, so we defer further investigation to future work.

\section{Discussion}

Mystique has successfully generated synthetic benchmarks for production AI workloads and has already been used in several challenging system studies, by both internal teams and external partners, such as Intel, NVIDIA, and AMD. At the same time, there are even more interesting directions we plan to explore with this new methodology.

\subsection{Data processing and AI tax}

The importance of considering the end-to-end behavior of AI applications is increasing due to the rise of data processing, also known as the AI tax~\cite{richins2020missing, buch2021ai, wu2022sustainable}. This is also our broader objective of generating AI benchmarks. As a first step, Mystique currently focuses on AI model components that are deployed on GPU servers, as 1) in many hyperscale deployment environments, including ours, data processing is deemed complex and critical, and as such, it is usually decoupled from the AI model, and assigned to dedicated CPU machines for execution, and 2) the execution of any processing left on the GPU server is often overshadowed by the well-tuned pipeline design, and requires minimal CPU resources compared to the massive consumption on the GPUs. Mystique is now better suited for training workloads than for inference workloads, where AI tax has a larger presence on the GPU server. It is our next step to utilize tools like Ditto~\cite{liang2023ditto} to incorporate the CPU execution fraction to our framework.

\subsection{Advanced ET analyzer and builder}

We are currently using a simple ET analyzer based on population weight to guide our selection of full ETs as replay samples from the trace database. This can be enhanced by incorporating more sophisticated weight calculations, such as timing costs. Moreover, we can explore operator-level summary and weighting to further improve the selection process, and leverage the composability of ET to combine portions from different ETs into a single replay trace for more efficient aggregation.


\subsection{Adaptability to other ML frameworks and domains}
\label{sec:support_other_frameworks}

The essence of Mystique is leveraging the appropriate level of abstraction provided by the framework to reproduce performance. Currently, our framework focuses on PyTorch models due to their widespread use in our environment, but it can be extended to other ML frameworks, such as TensorFlow, MXNet. These frameworks share a common feature, the use of computational graphs, such as ET in PyTorch, to express the model and execute operators. For example, TensorFlow provides the ability to save a graph using \texttt{tf.function} and to rerun it without the original code. Therefore, our cloning methodology is transferable to other frameworks, to create a synthetic benchmark that uses the same framework as the original application.

Furthermore, we are investigating the feasibility of generating benchmarks across different frameworks, e.g., replaying TensorFlow graphs using PyTorch, to maximize the use and impact of Mystique. To accomplish this, we are spearheading a collaboration between academic and industrial partners to propose a standard, portable format for execution traces that can be used across multiple frameworks.\footnote{https://github.com/chakra-et/chakra.} 

Beyond AI, we believe that with proper abstractions and profiling capabilities, Mystique can be adapted to other domains or domain-specific languages, for example, the RPC frameworks for cloud microservices.


\subsection{Improved privacy and IP protection}

Given that AI models can provide a significant competitive edge for a company, it is crucial to take extensive measures to protect them. However, this can create challenges when attempting to share the workload with external vendors for co-design or co-optimization. Mystique can potentially address this issue by obfuscating the real ETs and substituting important IP-protected blocks with performance-equivalent public ones. This approach allows for efficient sharing of the workload, while still capturing the intended performance behavior of a production workload.

\section{Conclusion}


We present Mystique, a new generation methodology for AI benchmarks, based on execution traces replay. 
Mystique addresses the scalability issues stemming from both the large model variety and the constantly changing workload landscape. We demonstrate that our methodology generates AI benchmarks highly similar to the original applications, while being easy to use and portable across platforms, without the need for regenerating. We have illustrated several use cases for Mystique, including early stage platform evaluation, subtrace replay, and scaled-down performance testing, all of which are highly challenging using existing techniques.

\begin{acks}
We sincerely thank David Berard and Valentin Andrei from Meta for their feedback to this work, and the anonymous reviewers for their suggestions on earlier versions of this manuscript. This work was in part sponsored by Meta through an internship and a student researcher appointment. This work was also in part supported by NSF CAREER Award CCF-1846046, an Intel Research Award, a
Sloan Research Fellowship, a Microsoft Research Fellowship, and a Facebook Research Faculty Award. 
\end{acks}

\balance

\bibliographystyle{ACM-Reference-Format}
\bibliography{refs}

\end{sloppypar}
\end{document}